# Identification and Characterisation of Technological Topics in the Field of Molecular Biology

Ivana Roche*

INIST-CNRS, 2 allée du Parc de Brabois, CS 10310, 54519 Vandoeuvre-les-Nancy, France, Tel.: +33 (0) 383504600, Fax: +33 (0) 383504650, ivana.roche@inist.fr

Dominique Besagni, Claire François

INIST-CNRS, 2 allée du Parc de Brabois, CS 10310, 54519 Vandoeuvre-les-Nancy, France, Tel.: +33 (0) 383504600, Fax: +33 (0) 383504650, dominique.besagni@inist.fr, claire.francois@inist.fr

Marianne Hörlesberger, Edgar Schiebel

Austrian Research Centers GmbH, Tech Gate Vienna, Donau-City-Straße 1, 1220 Wien, Austria, Tel.: +43 (0) 50 550-4524, Fax : +43 (0) 50 550-4500,
marianne.hoerlesberger@arcs.ac.at, edgar.schiebel@arcs.ac.at

* Corresponding author

## 1 Background

This paper focuses on methodological approaches for characterising the specific topics within a technological field based on scientific literature data. We introduce a diachronic clustering analysis approach and some bibliometric indicators. The results are visualised with the software-tool Stanalyst® [1]. We are applying our methods to the field "Molecular Biology". This field has grown a great deal in the last decade.

## 2 Problem / application

How can we identify and characterise important topics in a set of several thousand articles? Which technological aspects can be detected? Which of them are already established and which of them are new? How are the topics linked to each other? We are trying to answer these questions by applying our bibliometric analysis methods to a set of scientific literature data recorded in the bibliographic database PASCAL.

## 3 Methodology

The data for our study were extracted from PASCAL database in function of their classification categories and keywords.

The diffusion model identifies three categories of terms: the established terms, terms unusual in this topic, and cross section terms. We apply the indicator TFIDF (text frequency inverse document frequency) adapted to our research question and the GINI coefficient, a measure of statistical dispersion most prominently used as a measure of inequality of income distribution or inequality of wealth distribution.

The diachronic cluster analysis is realized with the help of a clustering tool applying first the axial K-means method to produce a non-hierarchical clustering algorithm based on the neuronal formalism of Kohonen's self-organizing maps and then a principal component analysis to map the obtained clusters. This tool is implemented in the information analysis platform Stanalyst®. Considering two successive time periods, our diachronic approach enables us to follow the field time evolution by analysing the two obtained cluster sets and maps.

The applied methods are linked together as the following table shows.

|  |  | *Diffusion Model* | | |
|---|---|---|---|---|
|  |  | Unusual Terms | Established Terms | Cross Section Terms |
| *Diachronic Cluster Analysis* | Clusters in the First Period |  |  |  |
|  | Clusters in the Second Period | new terms | terms with roots in the first period | terms with roots in the first period |

## 4 Outcome/findings/results

The diachronic cluster analysis is made for two periods and shows which topics of the second period have roots in the first one and which topics are new in the second period.

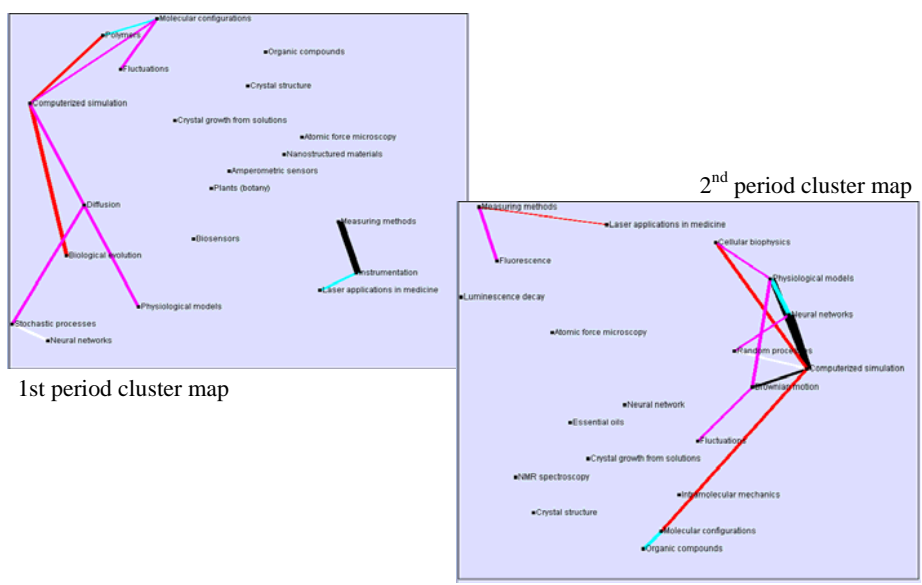

1st period cluster map

2nd period cluster map

On the map of both periods, as we can see in the above figure, it is possible to singularize an interesting dichotomous configuration represented by two very strongly connected cluster networks associating about two thirds of the clusters. However, these networks have quite different characteristics. On the one hand, the themes present in the biggest one are related to the modelling and simulation of biological phenomena. In the other hand, the little network is very homogeneous and deals essentially with instrumentation topics. The remaining clusters are scattered in the map with no significant links with the two above described networks.

Concerning the biggest network, located in the left side of the first period map and in the right side of the second period map, we can observe that structurally it remains the same with two sub-networks brought together by the cluster "Computerized simulation". But its content analysis allows us to find some interesting thematic evolutions. In the first period, its content is focused on both the description of physical characteristics of biological structures and the theoretical aspects related to the physiological process modelling using, for example, neural networks. In the second period, we can detect a refocusing of the couple formed by the clusters "Neural networks" and "Stochastic processes" and an impressive densification of the resulting sub-network with the new associated cluster "Brownian motion".

The smallest network, located in the right side of the first period map, deals with instrumentation techniques applied to measure and therapeutic issues. On the second period map, the network is located in the left side and shows a significant stability despite a reorganization of cluster contents.

## 5 Conclusion

"Molecular Biology" is a broad field. By applying the complementary methods presented here, it can be characterised and described presenting different views of the features of the field. On the other hand, the two methods point out the relationship of the different topics in our investigated field and their evolution.